# Information Theoretic Bounds for Low-Rank Matrix Completion

Sriram Vishwanath
Department of Electrical and Computer Engineering,
University of Texas at Austin
sriram@austin.utexas.edu

*Abstract*—This paper studies the low-rank matrix completion problem from an information theoretic perspective. The completion problem is rephrased as a communication problem of an (uncoded) low-rank matrix source over an erasure channel. The paper then uses achievability and converse arguments to present order-wise optimal bounds for the completion problem.

*Index Terms*—The Netflix prize

## I. Introduction

The low-rank matrix completion problem has been fairly well-studied in literature [3], [2], [1], [5], [6], with both algorithms for matrix completion and an analysis of the limits within which this is possible [4], [7]. In [4], the authors present optimality results quantifying the minimum number of entries needed to recover a matrix of rank $r$ (using any possible algorithm). Also, under certain incoherence assumptions on the singular vectors of the matrix, [4] shows that recovery is possible by solving a convenient convex program as soon as the number of entries is of the order of the bound (within polylog factors). The authors of [4] utilize a combination of multiple mathematical principles along with an optimization approach to determining these limits. In this paper, we study the low-rank matrix completion problem using a formulation similar to an information-theoretic coding problem and obtain achievability and converse bounds on near-perfect low-rank matrix completion that are similar to those in [4]. This reformulation of the low-rank matrix completion problem as a communication/compression problem enables us to generalize the near-perfect matrix completion problem to one which incorporates alternate models such as noise and distortion, and helps us gain insights into the connections between information-theoretic principles and matrix completion problems.

In [7], the authors show that to reconstruct a matrix of rank $r$ within an accuracy $\delta$, $C(r,\delta)n$ observations are sufficient. Results on low-rank matrix completion with noise are presented in [6] (and citations therein). The lossy matrix completion problem bears a strong resemblance to a quantization/rate distortion problem, while the low-rank matrix completion with noise problem has close intuitive connections with a channel coding problem. This paper is aimed at being a first step in making these connections more concrete.

This work is supported by NSF grants CCF-0934924, CCF-0916713 and CCF-0905200.

One of the intuitive connections between conventional information-theoretic coding theorems and the low-rank matrix completion problem is the "erasure-source-channel" perspective. The analogy can be drawn between the two as follows: Consider a system where the transmit source is fixed to be the set of all $m \times n$ matrices of rank $r$ or less. When the source is transmitted (in an uncoded fashion), the "communication channel" causes random erasures in $k$ positions of each transmitted matrix. The goal is to recover the original source with high probability at the receiver. The matrix completion problem is then rephrased as: how large can the number of random erasures $k$ be so that it is possible to distinguish each element of the matrix-source with high probability at the receiver? Although we do not explicitly use this reformulation of the matrix completion problem in our theorem statement or proofs, it is a useful analogy to remember as we proceed through the remainder of the paper.

Note that the low-rank matrix completion problem setting is in some ways different from conventional source and/or channel coding literature. For example, it does not incorporate an encoding process. The source (rank $r$ matrices) are directly transmitted and the channel is tightly coupled with the source. Regardless of differences, we endeavor in this paper to highlight the similarities and to point out that many of the existing tools in information and coding theory [8] may be directly applicable to addressing problems in the domain of matrix completion.

In summary, the main contributions of this paper are:

1) Bounds using tools from information-theory for low-rank matrix completion.
2) For an $m \times m$ matrix of rank $r$, a lower bound of $\Omega(m)$ and an achievable near-perfect reconstruction with $\Theta(m \log m)$ randomly chosen samples (for large $m$ and large alphabet size).
3) Lower bounds for matrix reconstruction with distortion constraints using concepts from rate-distortion.

The rest of this paper is organized as follows: the next section formally presents the system model. In Section III, we study both the achievability and converse bounds for the case of near-perfect matrix completion. In Section IV, we present lower-bounds for the case when we desire to learn low-rank matrices within a distortion constraint. We conclude the paper with Section V.

## II. SYSTEM MODEL

First, a note on the notation used in this paper: $\mathcal{S}$ denotes a set and $|\mathcal{S}|$ denotes the cardinality of the set $\mathcal{S}$. $Y^n$ denotes a vector of n-entries $Y_1, \ldots, Y_n$, while $Y_j^k$ for $j < k$ denotes the subvector $Y_j, \ldots, Y_k$. $S$ is used to denote both the random variable and a particular realization of it. $\Pr(.)$ denotes the probability of a certain event.

Let $\mathcal{S}$ be the set of all $m \times m$ matrices with the following structure:

$$S = UV \quad (1)$$

$\forall S \in \mathcal{S}$. Here, $U$ is an $m \times r$ matrix, and $V$ is an $r \times m$ matrix. The entries of $U$ and $V$ are assumed to belong to the finite field $\mathbb{Z}_q$, and the matrix multiplication is defined over integers $\mathbb{Z}$.

We make two assumptions on $S$ for the sake of simplicity: first, that it is of equal dimension $m \times m$. Second, that the entries of $U$ and $V$ are assumed to be drawn uniformly and independently from $\mathbb{Z}_q$. Both of these assumptions can be relaxed relatively easily. Making such assumptions helps us derive relatively uncomplicated expressions for the relationships between system parameters that resemble those in [4], [7]. The expressions would be considerably more involved for more general models.

Note that the set $\mathcal{S}$ contains matrices of rank $r$ or less. However, as the size of the alphabet ($q$) increases, the probability that $(u, v) \in (\mathcal{U}, \mathcal{V})$ has a rank less than $r$ diminishes.

For any $S \in \mathcal{S}$, we are given $n$ randomly chosen (without replacement) values from $S$. We use $Y^n$ to represent the values of the matrix $S$ at those locations. We denote the locations $(i, j)$, $1 \leq i, j \leq m$ that were sampled as the vector $Z^n$. From $Y^n$ and $Z^n$, we desire to recover $S$. For a given value of $n$ and a *recovery* function $\hat{S} = g(Y^n, Z^n)$, we define

$$P_e \triangleq Pr\left(\hat{S} \neq S | Y^n, Z^n\right)$$

As in conventional analysis, we consider the probability of error averaged over all $S \in \mathcal{S}$. In the case when we desire near-perfect recovery of the matrix $S$, we desire that $n$ be large enough such that there exists a decoding function $g$ with "small" $P_e$. This is similar to a lossless source-recovery problem setting. We refer to this as the near-perfect recovery as there is a finite (but arbitrarily small) probability that the recovery process will fail. This problem formulation is analyzed in further detail in Section III.

Alternatively, we may impose a distortion constraint on the recovery process

$$\mathbb{E}[d(s, \hat{s})] \leq D$$

where $d(s, \hat{s})$ is a suitable distortion function. Again, we desire that $l$ be large enough such that the reconstructed $\hat{S}$ meets this distortion requirement. This bears a strong resemblance to a rate distortion problem setting and lower bounds for it are studied in greater detail in Section IV.

In this paper, we determine the relationship between $m$ and $n$ that is required to recover $\hat{S}$ within the appropriate constraint for a given fixed rank $r$. We determine this relationship in the order sense when all of $m$, $n$ and alphabet size $q$ are sufficiently large.

## III. NEAR-PERFECT MATRIX RECOVERY

In this setting, we desire that, for any $\epsilon > 0$, there exist an $n$ and correspondingly, an $l$ sufficiently large such that, on an average across all elements of $\mathcal{S}$, the elements of $S$ can be recovered with a probability $P_e < \epsilon$.

*Theorem 3.1:* Given an n-length sampled sequence $Y^n$ and sampled locations $Z^n$, a matrix from $S$ can be reconstructed with high probability only if

$$n = \Omega(m).$$

Moreover, if $n = \Theta(m \log m)$, a reconstruction algorithm exists that will determine $S$ accurately with high probability. Specifically, given a target probability of error $\epsilon$ and a finite rank $r$, there exists an $m, q$ large enough and an $n = \Theta(m \log m)$ such that $P_e \leq \epsilon$.

**Proof:** In the same spirit as a channel-coding theorem, this proof incorporates both an achievability and a converse component. We begin with the converse argument:

### A. Converse

From Fano's inequality [8], we have that

$$H(S|Y^n, Z^n) \leq P_e \log |\mathcal{S}| \leq P_e(2rm \log q)$$

Therefore, we have:

$$
\begin{aligned}
H(S) &\stackrel{(a)}{=} H(S|Z^n) \\
&\stackrel{(b)}{\leq} I(S; Y^n | Z^n) + P_e(2rm \log q) \\
&\stackrel{(c)}{=} H(Y^n|Z^n) - H(Y^n|S, Z^n) + P_e(2rm \log q) \\
&\stackrel{(d)}{=} H(Y^n|Z^n) + P_e(2rm \log q) \\
&\stackrel{(e)}{\leq} n \log(rq^2) + P_e(2rm \log q)
\end{aligned}
$$

where (a) follows from the independence between $S$ and $Z^n$, (b) from Fano's inequality, (c) from the chain rule on mutual information and (d) from the fact that $Y^n$ is a deterministic function of $S$ given $Z^n$. Finally, (e) follows from the realization that each entry of any matrix $S \in \mathcal{S}$ has a maximum value of $rq^2$ (from the definition of $\mathcal{S}$ in Equation 1). So we have,

$$H(Y^n|Z^n) \leq \sum_i H(Y_i|Z_i) \leq n \log(rq^2)$$

But, we also have that

$$H(S) = H(UV) \geq H(UV|V) = H(U) = mr \log q$$

Thus we must have

$$mr \log q \leq n \log(rq^2) + P_e(2rm \log q)$$

So for $P_e$ arbitrarily small, an $n = \Omega(m)$ is necessary for reconstruction. □

Note that this can also be seen directly using a fairly intuitive and straightforward degrees-of-freedom argument for the system.

Next, we proceed to the achievability argument.

## B. Achievability

The achievability argument is the more involved component of this proof. Define $A_\epsilon^m(S)$ as the set of all $\epsilon$-typical matrices $S \in \mathcal{S}$ generated in accordance with (1). First, we define the sets [8]:

$$
\begin{aligned}
A_\epsilon^m(\mathcal{U}) &= \left\{ u \in \mathcal{U} : \left| -\frac{1}{rm} \log p(u) - \log q \right| \leq \epsilon \right\} \\
A_\epsilon^m(\mathcal{V}) &= \left\{ v \in \mathcal{V} : \left| -\frac{1}{rm} \log p(v) - \log q \right| \leq \epsilon \right\} \\
A_\epsilon^m(\mathcal{S}) &= \{ s = uv, u \in A_\epsilon^m(\mathcal{U}), v \in A_\epsilon^m(\mathcal{V}) \}
\end{aligned}
\qquad (2)
$$

Note that $|\mathcal{U}| = |\mathcal{V}| = 2^{rm \log q}$. Therefore, we have that:

$$|A_\epsilon^m(\mathcal{S})| \leq |\mathcal{S}| \leq 2^{2rm \log q}$$

Now, we sample the set $A_\epsilon^m(\mathcal{S})$, dropping $2^{2rm\delta}$ of its entries at random to generate the set $\mathcal{T}$. Thus, we have

$$|\mathcal{T}| \leq 2^{2rm(\log q - \delta)}$$

Now, given that the "received vector" $Y^n, Z^n$ resulted from a matrix $S \in \mathcal{S}$, we "decode" the sparse matrix as follows: we determine all $\hat{S} \in A_\epsilon^m$ that match the values $Y^n$ in the positions corresponding to $Z^n$. We declare success if a unique $\hat{S}$ is found, and declare an error if:

1) The event $E_0$ occurs, which is $S \notin \mathcal{T}$, or
2) The event $E_s'$ occurs - there exists $S' \neq S \in \mathcal{T}$ that agrees with $Y^n$ in the positions $Z^n$.

The overall probability of error is given by

$$P_e = Pr\left( E_0 \bigcup \bigcup_{S' \in \mathcal{T}, S' \neq S} E_S' \right)$$

It follows from AEP [8] that:

$$\Pr(\mathcal{T}) \geq 1 - \gamma(\delta) \qquad (3)$$

where $\gamma(\delta)$ goes to zero as $\delta \to 0$ and $m \to \infty$. Therefore,

$$P_e \leq \gamma(\delta) + \sum_{S' \in \mathcal{T}, S' \neq S} \Pr(E_S').$$

It is important to note that, for a particular value of $Y^n = y^n, Z^n = z^n$, either the event $E_s'$ occurs (with probability 1) or it does not occur at all. The key step here is to average this over all realizations of $Y^n$ and all possible sampling strategies $Z^n$.

To determine the remainder of $P_e$, we need the following two lemmas:

*Lemma 3.2:* Let $A = [a_1, a_2, \ldots a_r]$ be a random vector uniformly chosen over $\mathbb{Z}_q$, and let $C$ be an $r \times r$ random matrix with entries from $\mathbb{Z}_q$ with the $i$th column denoted as $C_i$. Then, for any $\beta > 0$, there exists a $q$ sufficiently large such that:

$$
\begin{aligned}
H(C_r A | C_1 A, C_2 A, \ldots, C_{r-1} A, C) &\geq \left(1 - \frac{r^2}{q}\right) \log q \\
&\geq \log q - \beta.
\end{aligned}
$$

Proof: Note that:

$$H(C_r A | C_1 A, \ldots, C_{r-1} A, C) = H(CA|C) - H(C_1 A, \ldots, C_{r-1} A | C)$$

As noted in [11], [12], the probability that $C$ is not invertible (for both integer-valued and finite-field matrices) diminishes at least as $r/q$ (Schwartz-Zippel lemma). Thus

$$H(CA|C) \geq \left(1 - \frac{r}{q}\right) r \log q$$

and

$$H(C_1 A, C_2 A, \ldots, C_{r-1} A | C) \leq (r-1) \log q$$

Thus we have the result. $\square$

Note that

$$H(C_r A | \tilde{C} A, C) \geq H(C_r A | C_1 A, C_2 A, \ldots, C_{r-1} A, C)$$

where $\tilde{C}$ is any subset of $C_1, C_2, \ldots, C_{r-1}$. Therefore, we must have:

$$H(C_r A | \tilde{C} A, C) \geq \log q - \beta$$

*Lemma 3.3:* For an arbitrary $\xi > 0$, there exists an $n_2, m_2$ such that, for $n > n_2$ and $m > m_2$, we have:

$$\Pr(E_s' | Z^n = z^n) \leq 2^{-(H(Y^n | Z^n = z^n) - \frac{n\xi}{\log n})}$$

Proof: Note that:

$$\Pr(\exists S' \in \mathcal{T} : S' \neq S | Z^n = z^n) = \Pr(Y^n | Z^n = z^n). \qquad (4)$$

In words, the probability that two distinct elements of $\mathcal{T}$ agree in a given set of $n$ randomly chosen places is equal to the probability of that particular set of values, across all possibilities when sampling matrices in $\mathcal{T}$.

The second part of the proof is essentially the Shannon-McMillan-Breiman Theorem (SMB) for discrete-time discrete-valued sources with minor modifications. As the proof of the SMB theorem is fairly involved, we refer the reader to the sandwich proof by Algoet and Cover [9], which is summarized in [10].

Next, we quantify $H(Y^n | Z^n = z^n)$. To do so, note that

$$H(Y_n | Y^{n-1}, z^n) \geq \max\{H(Y_n | Y^{n-1}, z^n, V), H(Y_n | Y^{n-1}, z^n, U)\}$$

which follows from the fact that conditioning cannot increase entropy. Next, we determine $H(Y_n | Y^{n-1}, Z^n = z^n, V)$ noting that an analogous exercise holds for $H(Y_n | Y^{n-1}, Z^n = z^n, U)$. Given $V$, $Y_n$ is a linear combination of the entries of row of $U$ using known coefficients from $V$ chosen through $Z_n = z_n$. Let this row be denoted as $U_i$. If $z^{n-1}$ causes $Y^{n-1}$ to contain $r - 1$ or less linear combinations of $U_i$, then from Lemma 3.2, we have that:

$$H(Y_n | Y^{n-1}, Z^n = z^n, V) \geq \log q - \beta.$$

Otherwise, we use the trivial lower bound $H(Y_n | Y^{n-1}, Z^n = z^n, V) \geq 0$.

A similar inequality holds for $H(Y_n|Y^{n-1}, Z^n = z^n, U)$ if $z^n$ causes $Y^{n-1}$ to have $r-1$ or less linear combinations of the particular column of $V$ in $Y_n$. In case we have $r$ or more linear combinations, we assign $H(Y_n|Y^{n-1}, Z^n = z^n, U) \geq 0$.

For the remainder of the achievability argument (Equation (7)), we desire that the number of samples $n$ be such that

$$H(Y^n|Z^n = z^n) \geq 2rm(\log q - \beta). \quad (5)$$

Note that the upper limit on $H(Y^n|Z^n = z^n)$ is $2rm \log q$, and thus (5) is "close" to this limit for small $\beta$. This may or may not hold, depending on $z^n$. We require that $n$ be large enough so that a "typical" $Z^n$ result in each row and column of the sampled matrix have at least $r$ entries.

Let $G_n$ denote the set of all sampling sequences $Z^n = z^n$ that include at least $r$ entries in each row and column. We designate a new error event to include as $z^n$s that do not incorporate this requirement. We show that $n = \Theta(m \log m)$ is sufficient to ensure that $G_n$ occurs with high probability. This problem resembles the scenario where we have $n$ balls and $m$ bins, each bin with a capacity limited to a total of $m$ balls. We place the $n$ balls uniformly randomly in the $m$ bins sequentially, eliminating bins that are at capacity. We desire that the probability of any bin having $r - 1$ or less balls be small.

In the analysis that follows, we drop the max capacity of $m$ per bin as it can only lead to a larger value for $n$ to satisfy the requirement that each bin have at least $r$ balls, and study the problem of placing $n$ balls randomly in $m$ bins. Let $n = \alpha m \log m$ for any $\alpha > 2$. Then we have that the average number of balls in each bin is $\alpha \log m$. If $W_i$ is the number of balls in Bin $i$, using a Chernoff bound we have:

$$\Pr(W_i < r) \leq e^{-\frac{\alpha \log m}{2}\left(1 - \frac{r}{\alpha \log m}\right)^2}$$
$$\leq \frac{1}{m^{\alpha/2}}$$

Hence, the probability than any row or column of the sampled matrix has fewer than $r$ entries is upper bounded by:

$$2m \Pr(W_i < r) \leq \frac{2m}{m^{\alpha/2}}$$

which diminishes as $m$ increases. Let $m_2$ be such that, for all $m \geq m_2$, we have

$$2m \Pr(W_i < r) \leq \tau \quad (6)$$

for an arbitrary $\tau > 0$. As mentioned before, we declare an error when the sampled matrix is such that there are fewer than $r$ entries in any row or column. Therefore, the overall probability of error expression can be upper bounded as:

$$P_e \leq \gamma(\delta) + \Pr(Z^n \notin G_n)$$
$$+ \Pr(Z^n \in G_n) 2^{-(H(Y^n|Z^n=z^n) - \frac{n\xi}{\log n})} |\mathcal{T}|$$

From (5) and (6), when $n = \alpha m \log m$ we have

$$P_e \leq \gamma(\delta) + \tau + 2^{-\left(2rm\delta - rm\beta - \alpha m \log m \frac{\xi}{\log m}\right)} \quad (7)$$

Thus, as long as we choose

$$\delta > \frac{\beta}{2} + \frac{\alpha \xi}{2r},$$

there exists an $m_3$ large enough such that, for all $m > m_4$ we have:

$$2^{-(2rm\delta - rm\beta - \alpha m\xi)} \leq \lambda$$

for some $\lambda > 0$. Thus for an $m$ large enough, $P_e \leq \epsilon$ for any $\epsilon > 0$. This concludes the achievability proof. □

Thus, the overall result is established. Note that there is a log factor gap between the lower and upper bounds on matrix completion. This log-factor ensures enough entries are sampled from each row and column of the matrix. If a more systematic sampling method $Z^n$ was adopted for obtaining $Y^n$ from the matrix $S$ than just random sampling, then this log-factor may not be essential for near-perfect reconstruction.

## IV. MATRIX RECONSTRUCTION UNDER DISTORTION CONSTRAINTS

Next, we present lower bounds when we do not desire perfect reconstruction but allow for a distortion between the original matrix source $S$ and the reconstruction $\hat{S}$ as given by (9). We base this lower bound on principles from rate-distortion theory. The achievability argument is fairly involved and is therefore relegated to a future paper. We provide the lower bound as it is relatively straightforward to obtain and it illustrates the application of concepts from rate distortion theory to matrix reconstruction.

In this section, we present lower bounds under two settings - when the alphabet is discrete (under Hamming distortion) and when it is continuous (under squared error distortion).

### A. Case 1: Discrete source with Hamming distortion

Here, we desire to determine a bound on $n$ such that

$$\sum 1[(\hat{S} \neq S)_{i,j}] \leq Dm^\beta$$

Intuitively, we desire that the matrices S and $\hat{S}$ differ in $D$ places on an average. To determine the lower bound, we have the following inequalities:

$$H(\hat{S}|Z^n) \stackrel{(a)}{=} H(\hat{S}|Z^n) - H(\hat{S}|Y^n, Z^n)$$
$$H(\hat{S}|Z^n) - H(\hat{S}|S, Z^n) \leq I(\hat{S}; Y^n|Z^n)$$
$$I(S; \hat{S}|Z^n) \leq H(Y^n|Z^n) - H(Y^n|\hat{S}, Z^n)$$
$$I(S; \hat{S}|Z^n) \stackrel{(b)}{\leq} H(Y^n|Z^n)$$

where $(a)$ follows from the fact that $\hat{S}$ is a function of $Y^n, Z^n$, and $(b)$ from the fact that $Y^n$ and $\hat{S}$ must agree on the positions given by $Z^n$ to minimize distortion.

Now we have:

$$I(S; \hat{S}|Z^n) = H(S) - H(S|\hat{S}, Z^n)$$
$$\geq H(S) - H(S - \hat{S})$$

If $T \triangleq S - \hat{S}$, the distortion constraint requires that $T$ be a matrix with at most $Dm^\beta$ non-zero values, with a range of at

most $-rq^2$ to $rq^2$. From the maximum entropy theorem, we have

$$H(T) \leq Dm^\beta \log(2rq^2)$$

and so,

$$\begin{aligned}I(S;\hat{S}) &\geq H(S) - Dm^\beta \log(2rq^2) \\ &\geq 2rm(\log q - \delta) - Dm^\beta \log(2rq^2)\end{aligned} \quad (8)$$

Combining (8) and realizing that $H(Y^n) \leq n \log rq^2$, we have

$$n \log(rq^2) \geq 2rm(\log q - \delta) - Dm^\beta \log(2rq^2) \quad (9)$$

*Remark 4.1:* Note that if $\beta \geq 1$, then the lower bound (9), if tight, indicates that lossy reconstruction may be possible with a constant or polylog number of samples. However, the lower bound may not be tight in that regime and an achievability argument is needed to indicate if this is possible.

### B. Case 2: Continuous source with the squared norm

To illustrate the usefulness of the information-theoretic formulation, we consider the problem of reconstructing a matrix from a continuous alphabet. In this case, the source is any continuous valued matrix source of rank $r$ with a finite (differential) entropy rate:

$$h^*(S) \triangleq \lim_{m \to \infty} \frac{h(S)}{rm}$$

our distortion constraint is given by

$$\mathbb{E}[\sum (S - \hat{S})^2_{i,j}] \leq Dm^\beta \quad (10)$$

By the data processing inequality,

$$I(S; \hat{S}|Z^n) \leq I(S; Y^n|Z^n)$$

$$\begin{aligned}I(S; \hat{S}|Z^n) &= h(S) - h(S|\hat{S}, Z^n) &(11)\\ &\geq h(S) - h(S - \hat{S}) &(12)\\ & &(13)\end{aligned}$$

let $E \triangleq S - \hat{S}$ denote the error matrix. What we desire is to determine the maximum entropy rate of $E$ such that the entries of $E$ satisfy (10). Thus, the optimization problem is as follows:

$$\max h(E)$$

such that

$$\mathbb{E} \sum_{i,j} (E_{ij}) \leq Dm^\beta$$

Let $f(E)$ denote the 'true' joint distribution of the entries of $E$, and let $g(E)$ be a Gaussian distribution over $m^\beta$ entries of $E_{ij}$ such that $E_{ij}$ are independent with mean zero and variance $\sigma^2$ given by $\sigma^2 = D$.

We pick the remainder of $E_{ij} \equiv 0$. Given these, we have:

$$D(f||g) = -h(f) + \sum_{ij} \mathbb{E} \frac{(E_{ij})^2}{2\sigma^2} + m^\beta \frac{1}{2} \log(2\pi D)$$

Note that $D(f||g) \geq 0$, and so

$$h(f) \leq \frac{m^\beta}{2} \log(2\pi e \sigma^2)$$

Substituting $\sigma^2$ into this expression, we get, for $\beta < 2$,

$$h(E) \leq \frac{m^\beta}{2} \log(2\pi e D)$$

and therefore the resulting bound on the rate distortion function is:

$$I(S; \hat{S}) \geq rmh^*(S) - m^\beta \log(2\pi e D)$$

*Remark 4.2:* Note that if the distortion constraint $D = 0$, then reconstruction is impossible unless $n = m^2$. It is also trivial to see that for $\beta \geq 2$, only a few samples are required asymptotically for reconstruction.

## V. CONCLUSION

In this paper, we consider an information-theoretic formulation of the low-rank matrix completion problem. By using this formulation, we derive lower bounds on matrix reconstruction, and an upper bound in the case of near-perfect reconstruction.

A point to note that this paper does not provide low-complexity mechanisms for matrix reconstruction as in [4], [6], [5], [7]. In spite of this, this connection with information theory proves useful in analyzing the limits of matrix reconstruction under different models and constraints.

## VI. ACKNOWLEDGMENT

The author thanks Shreeshankar Bodas, Sujay Sanghavi and Brian Smith for insightful discussions and comments.

## REFERENCES

[1] A. Singer and M. Cucuringu, "Uniqueness of Low-Rank Matrix Completion by Rigidity Theory", submitted 2009. arXiv:0902.3846.
[2] J-F. Cai, E. J. Candès, and Z. Shen, "A singular value thresholding algorithm for matrix completion", *Technical report*, 2008. arXiv:0810.3286.
[3] B. Recht, M. Fazel, and P. A. Parrilo, "Guaranteed minimum rank solutions to linear matrix equations via nuclear norm minimization", preprint (2007), submitted to SIAM Review. arXiv:0706.4138.
[4] E. J. Candès and T. Tao, "The power of convex relaxation: Near-optimal matrix completion", *IEEE Trans. Inform. Theory*, to appear. arXiv:0903.1476.
[5] M. Fazel, E. Cand'es, B. Recht, and P. Parrilo, "Compressed sensing and robust recovery of low rank matrices", *Proc. Asilomar Conference*, Pacific Grove, CA, October 2008.
[6] E. J. Candès and Y. Plan, "Matrix completion with noise,"*Proceedings of the IEEE*, to appear. arXiv:0903.3131.
[7] R. Keshavan, A. Montanari and S. Oh "Learning low rank matrices from O(n) entries ," Proc. Allerton Conference, 2008. arXiv:0812.2599.
[8] T. Cover and J. Thomas, "Elements of Information Theory", Wiley 1991.
[9] P. H. Algoet and T. Cover, "A Sandwich Proof of the Shannon-McMillan-Breiman Theorem", *The Annals of Probability*, Vol. 16, No. 2, pp. 899-909, 1988.
[10] http://en.wikipedia.org/wiki/Asymptotic_ equipartition_ property

[11] T. Ho, R. Koetter, M. Medard, D. R. Karger and M. Effros, "The Benefits of Coding over Routing in a Randomized Setting", *Proc. IEEE International Symposium on Information Theory*, 2003.
[12] J. Bourgain, V. Vu and P. Wood, "On the singularity probability of discrete random matrices", arXiv:0905.0461.